\documentclass[fleqn,usenatbib]{mnras}

\usepackage{newtxtext,newtxmath}

\usepackage{soul}

\usepackage[T1]{fontenc}

\DeclareRobustCommand{\VAN}[3]{#2}
\let\VANthebibliography\thebibliography
\def\thebibliography{\DeclareRobustCommand{\VAN}[3]{##3}\VANthebibliography}

\usepackage{graphicx}	
\usepackage{amsmath}	
\usepackage{amssymb}	
\usepackage{xspace}

\newcommand{\NICER}{{\em NICER}\xspace}
\newcommand{\RXTE}{{\em RXTE}\xspace}
\newcommand{\diskbb}{{\tt diskbb}\xspace}
\newcommand{\bb}{{\tt bbody}\xspace}
\newcommand{\vkompthdk}{{\tt vkompthdk}\xspace}
\newcommand{\vkdualdk}{{\tt vkdualdk}\xspace}

\newcommand{\vkompth}{{\tt vkompth}\xspace}

\newcommand{\vkompthbb}{{\tt vkompthbb}\xspace}

\newcommand{\MAXI}{{MAXI~J1348--630}\xspace}
\def\lessim{\mathbin{\lower 3pt\hbox
     {$\rlap{\raise 5pt\hbox{$\char'074$}}\mathchar"7218$}}}   

\title[vKompth: a variable Comptonisation model for QPOs]{vKompth: A variable Comptonisation model for low-frequency quasi-periodic oscillations in black-hole X-ray binaries}

\author[Bellavita et al.]{
Candela Bellavita$^{1}$\thanks{CIN Fellow, bellavitacandela@gmail.com},
Federico Garc\'ia$^{2,3}$,
Mariano M\'endez$^{3}$,
Konstantinos Karpouzas$^{3}$
\\
$^{1}$Facultad de Ciencias Astron\'omicas y Geof\'{\i}sicas, Universidad Nacional de La Plata, Paseo del Bosque, B1900FWA La Plata, Argentina\\
$^{2}$Instituto Argentino de Radioastronom\'ia (CCT La Plata, CONICET; CICPBA; UNLP), C.C.5, (1894) Villa Elisa, Buenos Aires, Argentina\\
$^{3}$  Kapteyn Astronomical Institute, University of Groningen, PO BOX 800, NL-9700 AV Groningen, the Netherlands \\
}

\date{Accepted XXX. Received YYY; in original form ZZZ}

\pubyear{2022}

\begin{document}
\label{firstpage}
\pagerange{\pageref{firstpage}--\pageref{lastpage}}
\maketitle

\begin{abstract}
Low mass X-ray binaries (LMXBs) show strong variability over a broad range of time scales. The analysis of this variability, in particular of the quasi-periodic oscillations (QPO), is key to understanding the properties of the innermost regions of the accretion flow in these systems. We present a time-dependent Comptonisation model that fits the energy-dependent rms-amplitude and phase-lag spectra of low-frequency QPOs in black-hole (BH) LMXBs. We model the accretion disc as a multi-temperature blackbody source emitting soft photons which are then Compton up-scattered in a spherical corona, including feedback of Comptonised photons that return to the disc. We compare our results with those obtained with a model in which the seed-photons source is a spherical blackbody: at low energies the time-averaged, rms and phase-lag spectra are smoother for the disk-blackbody than for a blackbody, while at high energies both models give similar spectra. In general, we find that the rms increases with energy, the slope of the phase-lag spectrum depends strongly on the feedback, while the minimum-lag energy is correlated with the disc temperature. We fit the model to a 4.45-Hz type-B QPO in the BH LMXB MAXI J1438--630 and find statistically-better fits and more compatible parameters with the steady-state spectrum than those obtained with a blackbody seed-photons source. Furthermore, we successfully apply the model to the type-C QPO in the BH LMXB GRS 1915+105, and thus conclude that this variable-Comptonisation model reproduces the rms and phase-lags of both type B and C low-frequency QPOs in BH LMXBs.
\end{abstract}




\section{Introduction}

Low mass X-ray binaries (LMXB) are binary systems consisting of a compact object, a neutron star (NS) or a black hole (BH), accreting matter from a low-mass star. These sources present very high variability in their X-ray light curves over a broad range of time scales, from tens of milliseconds to years \citep{2016ASSL..440...61B,2021ASSL..461..263M}. The analysis of the variability is important because it gives us information about the accretion process in these objects \citep{1994ApJS...92..511V,1998ApJ...492L..59S,2009MNRAS.397L.101I,Kara,2019MNRAS.488..348M,ReigKylafisJet,2021ApJ...910L...3W}. 

Some BH-LMXBs are transient sources that during outburst describe a $q$ shape in the hardness-intensity diagram (HID) \citep{1997ApJ...479..926M,2001ApJS..132..377H,Belloni}. We can identify different states in the HID during an outburst of a BH-LMXB: at the start of the outburst the source is in the low hard state (LHS), with a hard energy spectrum. As the outburst progresses, the source follows an almost vertical path in the HID, increasing its intensity and barely decreasing its hardness. Eventually, the source transitions to the hard-intermediate and soft-intermediate states (HIMS and SIMS respectively) where, at an essentially constant intensity, the spectrum softens until the source reaches the high soft state (HSS). In this state the energy spectrum is soft and the intensity decreases while the hardness remains more or less constant. Finally, the source returns to the LHS completing the $q$ shape in the HID.

The power density spectra (PDS) of LMXBs present narrow features at characteristic frequencies, called quasi-periodic oscillations (QPOs), which correspond to variability in the X-ray light curve on well-defined timescales, providing information about the innermost regions of the accretion flow \citep{2020arXiv200108758I}. A QPO is characterized by a centroid frequency, $\nu_0$, a full-width half maximum, $\mathrm{FWHM}$, an amplitude, and the energy-dependent phase lag. 
Low-frequency QPOs (LF QPOs) observed in BH-LMXBs can be classified into three types, A, B, and C, according to their strength, quality factor, $Q= (\nu_0/\mathrm{FWHM})$, fractional root mean square (rms) amplitude, the properties of the underlying broadband noise in the PDS, and the spectral state of the source \citep{Belloni, Casella}. 
Type-A LF QPOs are broad peaks with centroid frequencies around 8~Hz and weak variability; they have a quality factor $Q<3$ and appear in the HSS. 
Type-B QPOs are narrower than type-A QPOs and have centroid frequencies around 6~Hz and a $Q>6$; they display a relatively weak variability, $\approx 4\%$ rms, and are found in the soft-intermediate state in the HID, where the spectrum is dominated by the soft component. 
Type C QPOs are narrow and strong features that peak around a variable centroid frequency between 0.1-15~Hz; they have a $Q>10$ and are very frequent in the hard and hard-intermediate states, where the spectrum is dominated by a hard component.

In this paper we propose a model that explains the rms and lag spectra of the LF-QPOs in BH-LMXB. While in principle the model that we describe here would be applicable to the three types of QPOs, type-A QPOs are very weak and up to date there is no report of an rms or lag spectrum of this type of QPOs. Hence, we will focus on type-B and C QPOs.

The X-ray spectrum of a BH-LMXBs is usually fitted with three additive components: a soft component emitted by an optically thick, geometrically thin accretion disk \citep{ShakuraSunyaev}, which dominates in the HSS, a hard component, produced in a plasma of highly energetic electrons \citep{SunyaevTitarchuk}, called the corona, that dominates in the LHS, and a reflection component \citep{Fabian}. The hard component is likely due to Comptonisation, which is the predominant interaction process between photons and hot electrons since it reproduces correctly the shape of the spectrum of X-ray sources at high energies \citep{1979Natur.279..506S,1997ApJ...480..735K}. The corona may well be a component of the accretion flow \citep{2009MNRAS.397L.101I} or the base of an out-flowing jet \citep{Giannios,Markoff, ReigKylafisJet}

For most QPOs in LMXBs, the variability increases with energy, reaching 20$-$30\% around 30~keV (eg.: for kHz QPOs \citealt{berger,Mendez2001,Gilfanov}, for HFQPOs in BHC \citealt{1997ApJ...482..993M,2013MNRAS.432...10B}, for type-C QPO \citealt{2002A&A...386..271R,2016ApJ...833...27Y,2017ApJ...845..143Z,2020MNRAS.494.1375Z}). At these energies, the contribution of the soft component to the X-ray spectrum is negligible; hence, the accretion disk alone cannot explain the rms amplitude behaviour. The recent measurements of a QPO up to $\sim$200~keV in MAXI~J1820+070 \citep{2020NatAs.tmp..184M} shows that reflection cannot explain the rms spectrum of this QPO because, even though reflection peaks around 20~keV, it decays at higher energies and it is unable to produce the variability observed up to 150-300~keV. Thus, the only possible explanation of the high rms amplitude of QPOs at (very) high energies is that the corona is leading the variability, either producing it directly or amplifying the oscillations produced in the accretion disk at different energy bands \citep{2006MNRAS.370..405S}.

Inverse Compton scattering naturally leads to a hard lag (the hard photons lag behind the soft ones) because the travel path inside the corona and the energy of a photon increase each time the photon is inverse-Compton scattered; because of this high-energy photons that suffered many scatterings escape the corona after low-energy photons that suffered less or no scatterings \citep{1988Natur.336..450M}. Some up-scattered photons may return to the disk and be re-emitted later. This mechanism, called {\em feedback}, produces a soft lag, where the less energetic photons are those that arrive later \citep{LeeMiller}.

Several papers proposed that the radiative properties of the variability in NS-LMXBs are due to the oscillation of thermodynamic properties of the corona or of the soft photon source \citep{LeeMiller, LeeMisra, KumarMisra14}. Recently, \citet{Karpouzas} presented a numerical time-dependent Comptonisation model that explains the energy-dependent rms and phase-lag spectrum of kiloHertz (kHz) QPOs in NS-LMXBs. Although this model was originally built to explain the kilohertz QPOs in NS LMXB, \citet{KostasTypeC}, \citet{2022mariano} and \citet{2022garcia} successfully applied the same model to the type-C LF-QPOs in the BH candidate GRS~1915+105. Furthermore, \citet{Garcia} used the same model to explain the type-B QPO of the BH MAXI J1348–630. In order to get a good fit of the phase lag and rms spectra, \citet{Garcia} considered two independent, but physically coupled, Comptonisation regions.

The results of \citet{LeeMisra}, \citet{KumarMisra14} and \citet{Karpouzas}, in the case of kilohertz QPOs in NS-LMXBs show that the rms and lag spectra are produced in a rather small, less than 10~km, corona illuminated by the spherical NS surface; it is therefore reasonable to model the emitting source as a spherical single-temperature blackbody as done by \citet{Karpouzas}. In a BH-LMXB, this assumption is no longer appropriate. In this case the source of seed photons is the accretion disk, and the corona is generally more extended than in NS-LMXBs \citep{Garcia,KostasTypeC,2022mariano,2022garcia}. Because of this, and based on \citet{Karpouzas}, we construct a model in which the soft photon spectrum is a multi-temperature blackbody \citep[\diskbb,][]{mitsuda1984}, to take into account that the emitting source is now the accretion disk. As in \citet{Karpouzas}, we continue assuming that the corona is spherically symmetric. 

The remainder of this paper is organized as follows:
In section \ref{model} we briefly describe the model of \citet{Karpouzas}, initially proposed to explain kHz QPOs in NS-LMXBs, and the changes that we introduced to explain the low frequencies QPOs in BH-LMXBs. In section \ref{results} we compare the predictions of our model with a disk as the seed spectrum to those of \citet{Karpouzas} black-body seed spectrum. In this section we also explore the behaviour of the rms and phase lag spectra for different values of the physical parameters involved. Subsequently, we use our model to fit the rms-amplitude and phase-lag spectra of the type-B QPO in \MAXI and the type-C QPO in GRS~1915+105 and compare our results with the ones obtained in \citet{Garcia} and \citet{2022garcia}, respectively. Finally, we discuss our results in Section~\ref{discussion}.

\section{The model} \label{model}

The model presented in \citet{Karpouzas} assumes that the surface of the NS, characterized by a radius $a$ and a temperature $T_{\rm s}$, injects photons into a spherical corona at a rate $\dot{n}_{\rm s \gamma}$ per unit volume. Since in this work we are interested in BH-LMXBs, we consider that the soft photon source is a geometrically thin and optically thick accretion disk, which is determined by the temperature, $T_{\rm s}$, at the innermost radius of the disk, $a$.

We model the corona as a spherically symmetric homogeneous plasma of size $L$, and temperature $T_{\rm e}$, filled with highly energetic electrons uniformly distributed with a number density $n_{\rm e}$. We assume that the seed photons are emitted as in \citet{SunyaevTitarchuk} and then inverse-Compton (IC) scattered in the corona, escaping at a rate $\dot{n}_{\rm esc}$ per unit volume. As \citet{Karpouzas}, we also consider feedback, by which a fraction $\eta$ of the photons scattered in the corona return to the soft photon source. Under these hypotheses, we are able to describe the evolution of the time-dependent photon spectrum, $n_{\gamma}(E,t)$, via the Kompaneets equation \citep{kompaneets}, which takes into account the multiple IC scatterings: 
\begin{equation}
\begin{split}
t_{\rm c} \frac{\partial{n_{\gamma}}} { \partial{t} } = \frac{1}{m_{\rm e} c^2} \frac{ \partial{} }{ \partial{E} } \Big( -4kT_{\rm e}En_{\gamma} + E^2 n_{\gamma} +kT_{\rm e}\frac{ \partial{} }{ \partial{E} }(E^2n_{\gamma}) \Big) \\ + t_{\rm c}\dot{n}_{\rm s
\gamma} - t_{\rm c}
\dot{n}_{\rm esc} \ \ ,
\label{eq:Kompaneets}
\end{split}
\end{equation}
where $m_{\rm e}$ is the electron rest mass, $c$ is the speed of light in vacuum, $k$ is the Boltzmann constant, and $t_{\rm c} = L/c \tau_{\rm T}$ is the Thomson collision time scale with $ \tau_{\rm T}$ the Thompson optical depth that, under our assumptions, it is given by $\tau_{\rm T} = \sigma_{\rm T} n_{\rm e} L$, where $\sigma_{\rm T}$ is the Thompson cross section and $n_{\rm e}$ the electron number density of the plasma. 

\citet{Karpouzas} assumed that the NS surface emits as a blackbody (\bb), so that the injection rate is
\begin{equation}
\begin{split}
\dot{n}_{\rm s \gamma} =  \Big[ \frac{3 a^2}{ [ (a+L)^3 - a^3 ] } \Big] \Big( \frac{2 \pi}{h^3c^2} {\rm BB}(E, kT_{\rm s}) \Big) \ \ , \\
\label{eq2:seed}
\end{split}
\end{equation}
where ${\rm BB}(E, kT_{\rm s}) =E^2 (e^{\frac{E}{kT_{\rm s}}}-1)^{-1} $. On the other hand, since we consider that in the case of a BH LMXB the soft photon source is the accretion disk, we model the emission with a multi-temperature disk blackbody  \citep[\diskbb,][]{mitsuda1984}, with inner radius $a$ and temperature at the inner radius $kT_s$.
The integrated flux of this model is equal to the one produced by a spherical \bb with radius $a$ and temperature $kT_s$. Thus, we can still use the $\dot{n}_{\rm s \gamma}$ in Eq.~(\ref{eq2:seed}) replacing ${\rm BB}(E, kT_s)$ by
${\rm DiskBB}(E,kT_s)$.

The photon escape rate per unit volume is:
\begin{equation}
\dot{n}_{\rm esc}=\frac{c \tau_{\rm T} n_{\gamma}(t,E)}{L N_{\rm esc} } = \frac{n_{\gamma}(t,E)}{t_{\rm c} N_{\rm esc}},
\label{eq:nesc}
\end{equation}
where $N_{\rm esc}$ is the number of scatterings that a photon undergoes before escaping from the corona and calculated as in \citet{Nesc}, assuming that each scattering is an independent event.

As in \citet{Karpouzas}, we model the QPO as a small oscillation of the spectrum around the time-averaged one, $n_{\gamma 0}$, obtained by solving the steady-state solution (SSS), i.e., imposing $ \frac{\partial n_{\gamma 0}}{\partial t} = 0 $ in Eq.~\ref{eq:Kompaneets}. This perturbation can be produced by the variation of any physical quantity involved in the Kompaneets equation and it can be written as $n_{\gamma} = n_{\gamma 0}(1 + \delta n_{\gamma} e^{-\omega t})$, where $\omega$ and $\delta n_{\gamma}$ are the QPO frequency and energy-dependent complex fractional amplitude, respectively. The modulus of $\delta n_{\gamma}$, $|\delta n_{\gamma}|$, is the rms amplitude of the QPO, while the argument of $\delta n_{\gamma}$, $\tan^{-1}\left(\frac{Im(\delta n_{\gamma} )}{Re(\delta n_{\gamma})}\right)$, is the QPO phase lag.

We assume that fluctuations of the spectrum are caused by perturbations in the coronal temperature, $T_{\rm e} = T_{\rm e 0}(1 + \delta T_{\rm e} e^{-\omega t})$, as a result of an oscillating external heating rate, $\dot{H}_{\rm ext} = \dot{H}_{\rm ext 0} (1 + \delta \dot{H}_{\rm ext} e^{-\omega t})$. The variation of the external heating rate, $\dot{H}_{\rm ext}$, allows us to explain the fact that,  after repeated scatterings, the temperature of the plasma does not drop significantly. Since we are also considering feedback, we assume that the temperature of the soft photon source oscillates due to the up-scattered photons that return to the disk, $T_{\rm s} = T_{\rm s 0}(1 + \delta T_{\rm s} e^{-\omega t})$. All three quantities, $\delta T_{\rm e}$, $\delta \dot{H}_{\rm ext}$, and $\delta T_{\rm s}$ are dimensionless complex quantities that can be calculated as explained in \citet{Karpouzas}.

\subsection{Solving scheme} \label{cuentas}
\label{sec:maths} 

In this section, we summarize the main steps implemented to solve Eq.~(\ref{eq:Kompaneets}).
First we calculate the SSS and use it to obtain $\delta n_{\gamma}$. We also define the new variables $x=E/kT_{\rm e}$ and $N = n_{\gamma}/n_{\rm c}$, where $n_{\rm c}=\frac{2 \pi m_{\rm e} t_{\rm c} k T_{\rm e0} 3 a^2}{h^3 [(a+L)^3-a^3]}$, and introduce them into Eq.~(\ref{eq:Kompaneets}) in order to work with a dimensionless equation.

In \citet{Karpouzas}, the numerical solution, $\delta{n_{\gamma}}$, was calculated using a linearly spaced energy grid. However, since the photon energy range of interest covers several orders of magnitude, for numerical purposes we use a logarithmically spaced energy grid instead. This change introduced to the solving scheme leads to a significant reduction of the computational time of the code, which is proportional to $M^2$, with $M$ the number of points in the mesh. We obtain a reasonable sample of the energy grid --even better than in the linear case at low energies, at which we are mostly interested-- using 10 times less points than with the linear grid; this reduce the computational time by a factor of 100. To this aim, we introduce the change of variable $u=\log(x)$ in the linearised form of Eq.~(\ref{eq:Kompaneets}).

The equation describing the SSS, after applying the change of variables mentioned above, and using finite differences in the linearized form of Eq~{\ref{eq:Kompaneets}}, results in 
\begin{equation}
\begin{split}
N^{j-1} \underbrace{ \left( \frac{1}{\delta u^2}-\frac{1}{2\delta u} \left( e^u-1\right) \right) }_{L(u_j)} + N^j \underbrace{ \left(  -\frac{2}{\delta u^2} -2  + 2 e^{u_j} - c_2 \right)}_{D(u_j)} \\ + N^{j+1} \underbrace{\left(  \frac{1}{\delta u^2} + \frac{1}{2\delta u} \left( e^u-1 \right) \right)}_{U(u_j)} = \underbrace{ - {\rm DiskBB}(e^{u_j}, kT_s) ~ e^{2u_j}}_{C(u_j)} ~ , 
\label{eq:dSS}
\end{split}
\end{equation}
where $j = 1, 2 ,..., M-1$ and $c_2 = \frac{m_{\rm e}c^2}{k T_{\rm e0} N_{\rm esc}(e^u)}$. We solve this equation as a boundary value problem with the boundary conditions $N^0=N^M=0$.

Once we find the array $N$ (one value at each energy in the grid), which is equivalent to $n_{\gamma 0}$, we obtain the following perturbed equation and calculate $\delta n_{\gamma}$ by proceeding similarly as in the case of the SSS:
\begin{equation}
{\scriptsize
\begin{split}
\delta n_{\gamma}^{j-1} \underbrace{ \Big( -\frac{1}{\delta u^2} + \frac{1}{2 \delta u} (e^u-1) + \frac{dN}{du} \Bigg|_{u_j}\frac{1}{N \delta u}  \Big) }_{L(u_j)}  \\ 
+\delta n_{\gamma} ^j \underbrace{\Big( \frac{2}{\delta u^2} + \frac{1}{N} \frac{{\rm DiskBB}(e^{u_j},kT_{\rm s0})}{(kT_{\rm e0})^2} - ic_5\Big)}_{D(u_j)} \\ + \delta n_{\gamma}^{j+1} \underbrace{\Big( -\frac{1}{\delta u_j^2} - \frac{1}{2 \delta u}(e^u-1) - \frac{dN}{du}\Bigg|_{u_j}\frac{1}{N \delta u}  \Big)}_{U(u_j)}  = \\
\underbrace{ \delta T_{\rm e} \Big(  - 2 +\frac{1}{N}\frac{d^2N}{du^2} \Bigg|_{u_j} + \frac{1}{N}\frac{dN}{du} \Bigg|_{u_j} \Big) + \delta T_{\rm s} \frac{1}{N} \frac{kT_{\rm s0}}{(kT_{\rm e0})^2} \frac{\partial {\rm DiskBB}(e^{u_j},kT_{\rm s})}{\partial T_{\rm s}}\Bigg|_{T_{\rm s0}}}_{C(u_j)},
\label{eq:dCODE}
\end{split}
}
\end{equation}
where the time derivative of ${\rm DiskBB}(e^u, kT_{\rm s})$ must be calculated numerically because, unlike the case of the \bb, the analytical expression is rather complicated given that the model involves an integration.

\section{Results} \label{results}

\begin{figure}
    \includegraphics[width=\columnwidth]{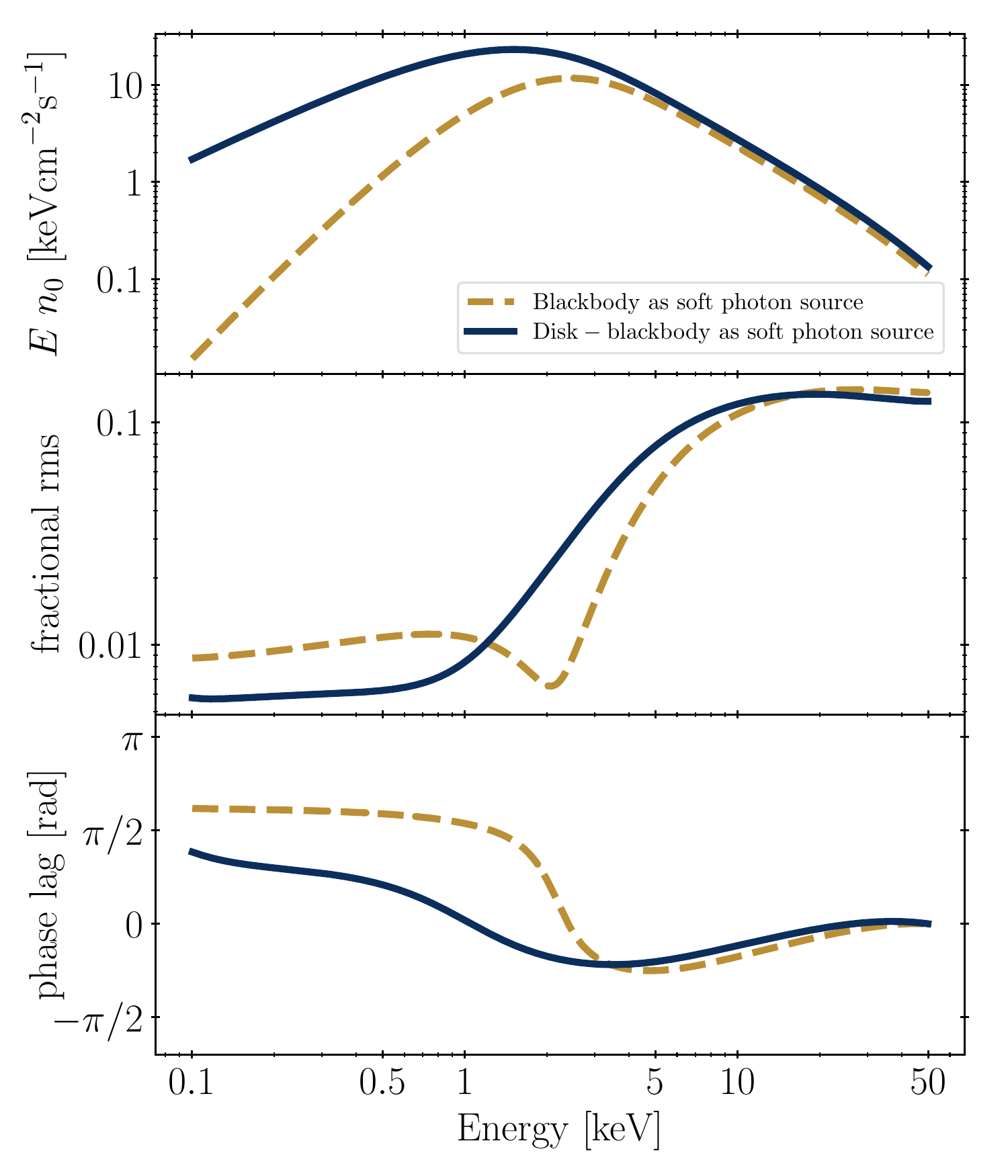}
    \caption{Top panel: The steady-state (i.e. time-averaged) spectra obtained from the solution using either a \bb (olive) or a \diskbb (blue) source of seed-soft photons. In the middle and lower panels we plot the energy dependent fractional rms amplitude and phase lag for each seed photon model (the colour convention is the same as in the top panel). In both cases we considered the following parameter values: $\eta=0.5$, $kT_s=0.5$~keV, $kT_e=20$~keV, $\Gamma=3.5$}
    \label{fig:BBvsDISK}
\end{figure}

In Fig.~\ref{fig:BBvsDISK}, we plot the SSS, the fractional rms amplitude, and the phase lags considering both soft photon sources, the \bb~(in olive dashed line) and the \diskbb (in blue solid line), using the same parameter values. We notice that both SSSs present a maximum around $1-3$~keV, behave in the same way at higher energies, but the \diskbb~is smoother than the \bb~at lower energies. Due to the pronounced peak in the SSS, the rms spectrum of the \bb~ shows a pivot point at the same energy of the maximum in the SSS, which is correlated with the abrupt drop of the lags. At lower energies, the \diskbb~rms spectrum shows less variability than that of the \bb. Consistently with the smoother behaviour of the SSS, the rms spectrum for the \diskbb~does not have a sharp minimum as in the case of the \bb. In the same way, we see that the drop of the lags for the \diskbb~is less pronounced than in the case of the \bb. At high energies, where all photons have already been Comptonised, both spectra become very similar.

In the following figures we present the fractional rms and phase lags obtained with our model, using a \diskbb~spectrum as the source of soft photons. We do so by independently varying each of the input parameters of the model. In all these figures, we use the same spectrum of Fig.~\ref{fig:BBvsDISK} as a guideline, which we plot with a solid black line.

\begin{figure*}
    \includegraphics[width=\columnwidth]{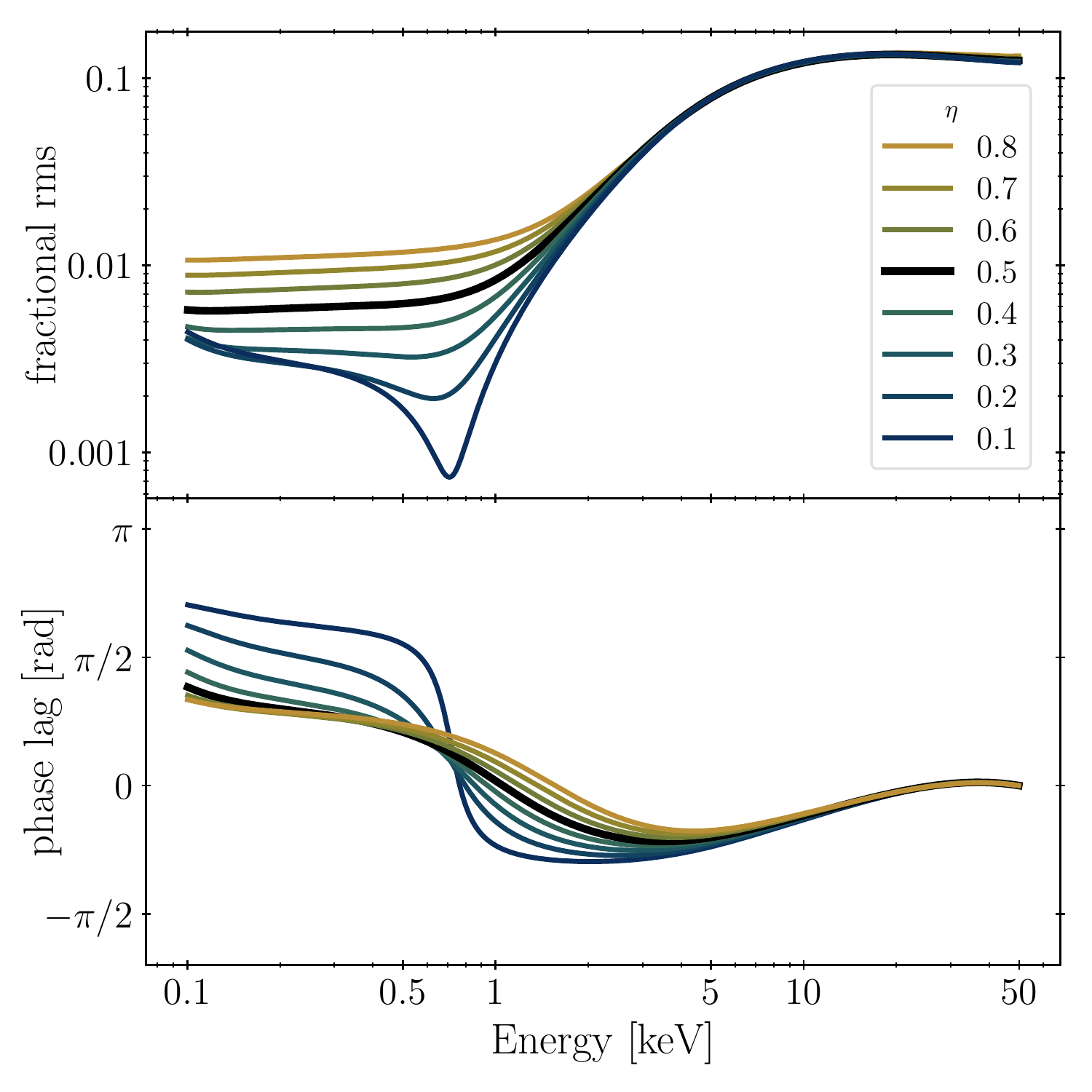} \,\,
    \includegraphics[width=\columnwidth]{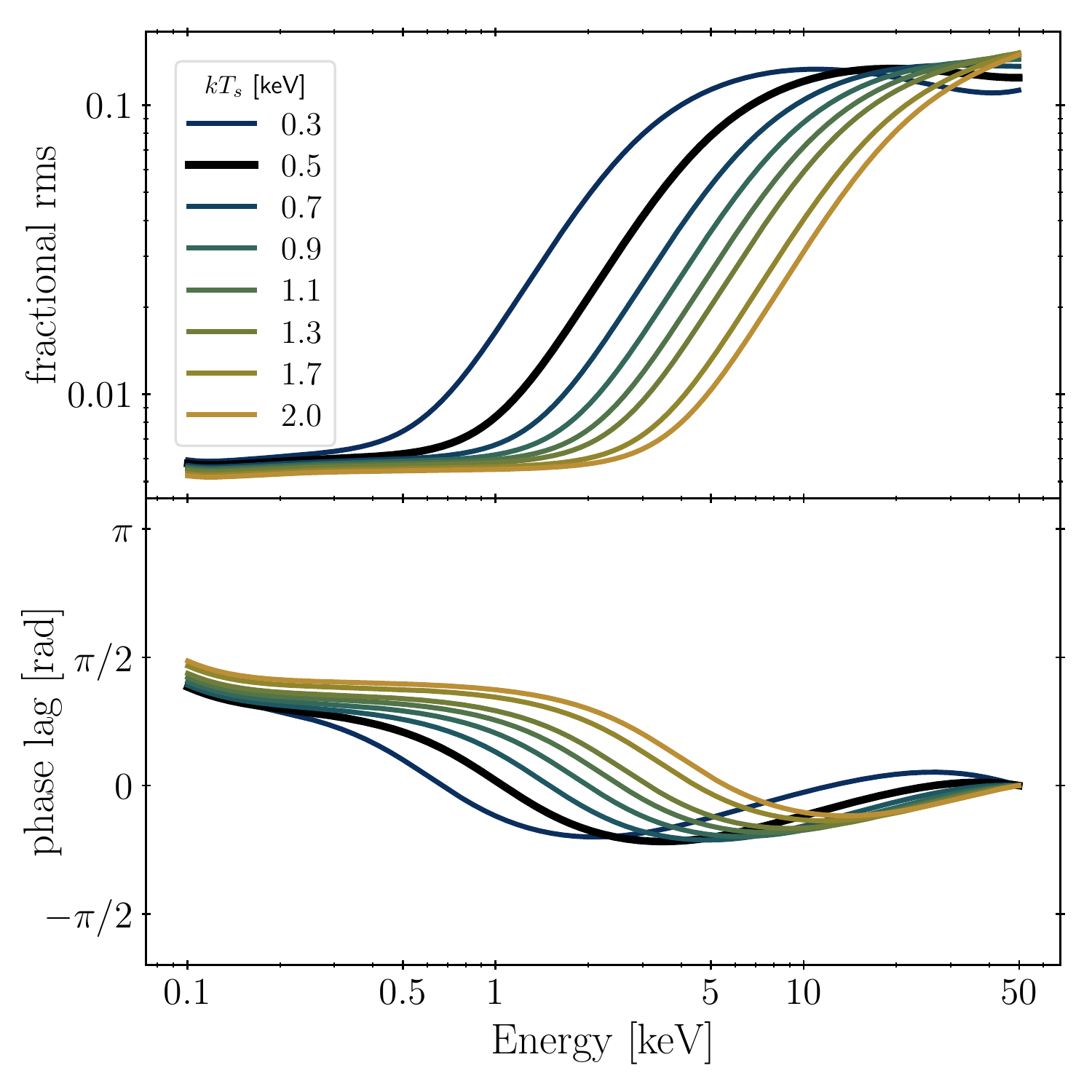}
    \caption{Plot of the energy dependent fractional rms and lags considering a disk-blackbody as soft photon source for different parameter values. On the left panel we show the behaviour of both spectra for different feedback values, from 0.1 (blue) to 0.8 (olive), leaving the other parameters fixed on $kT_s=0.5$~keV, $kT_e=20$~keV, $\Gamma=3.5$. On the right panel we fix the feedback to $\eta=0.5$ and we varied $kT_s$ to explore the spectra for different values of the soft photon source temperature, from 0.3~keV (blue) to 2.0~keV (olive). In both panels we display in black the same fractional rms and lag spectra as in Fig. \ref{fig:BBvsDISK}, using $\eta=0.5$, $kT_s=0.5$~keV, $kT_e=20$~keV, $\Gamma=3.5$ for a disk-blackbody case.}
    \label{fig:feedback_and_kTs}
\end{figure*}

\subsection{The feedback fraction and the soft-photon source}

On the left panel of Fig.~\ref{fig:feedback_and_kTs} we present the rms and lag spectra for different values of the feedback fraction parameter, $\eta$, ranging from very-low feedback, 0.1 in blue, to very-high feedback, 0.8 in olive. For this configuration, at high energies ($E \gtrsim 5$~keV) the model predicts that both the rms and lags are independent of the feedback fraction, showing essentially identical curves for both the rms and lags. This independence arises as feedback photons are re-emitted following the spectrum of the soft photon source, which dominates below 2~keV but is negligible above 5~keV. At low energies ($E \la 5$~keV) the curves start to separate according to the feedback fraction. For $\eta \gtrsim 0.4$, the rms amplitudes flatten to values of $\sim$1\% below 1~keV, while for $\eta < 0.4$ the rms decreases further around 0.8~keV, revealing a pivot point for the lowest values of the feedback fraction, and increases again to $\sim$0.5--0.8\% at 0.1~keV. At the same time, the lags have a minimum in the 1--5~keV energy range, with the energy of the minimum decreasing as the feedback fraction decreases. For energies below the pivot point ($E < 1$~keV), the lags increase again being sharper for the lowest feedback, as in the \bb~case (Garc\'ia et al., in prep.). 

On the right panel of Fig.~\ref{fig:feedback_and_kTs} we display the behaviour of the variability spectra for different values of the soft photon source temperature, $kT_s$, ranging from 0.4~keV in blue to 2~keV in olive. This figure shows that the model predicts essentially the same rms curves for all values of $kT_s$ but shifted to higher energies as the soft-photon temperature increases. For $kT_{\rm s} = 0.3$~keV the rise of the rms amplitude takes place at 0.4~keV and it moves to the right as $kT_{\rm s}$ increases, reaching $\sim$3~keV for a temperature of 2.0~keV. Similarly, the energy of the minimum in the lags is higher for higher values of $kT_s$, going from $\sim$2~keV for $kT_{\rm s} = 0.3$~keV to $\sim$14~keV for $kT_{\rm s}=2.0$~keV, suggesting a correlation between both quantities. We fitted the relation between $kT_{\rm s}$ and the energy of the minimum in the lag spectrum, $E_{\rm min}$, with a linear function and found a very significant correlation (correlation coefficient $r=0.999$, with a probability of $10^{-32}$ for the hypothesis that the data are not correlated), $E_{min} \approx 6.87 kT_s$ in keV.

\begin{figure*}
    \includegraphics[width=\columnwidth]{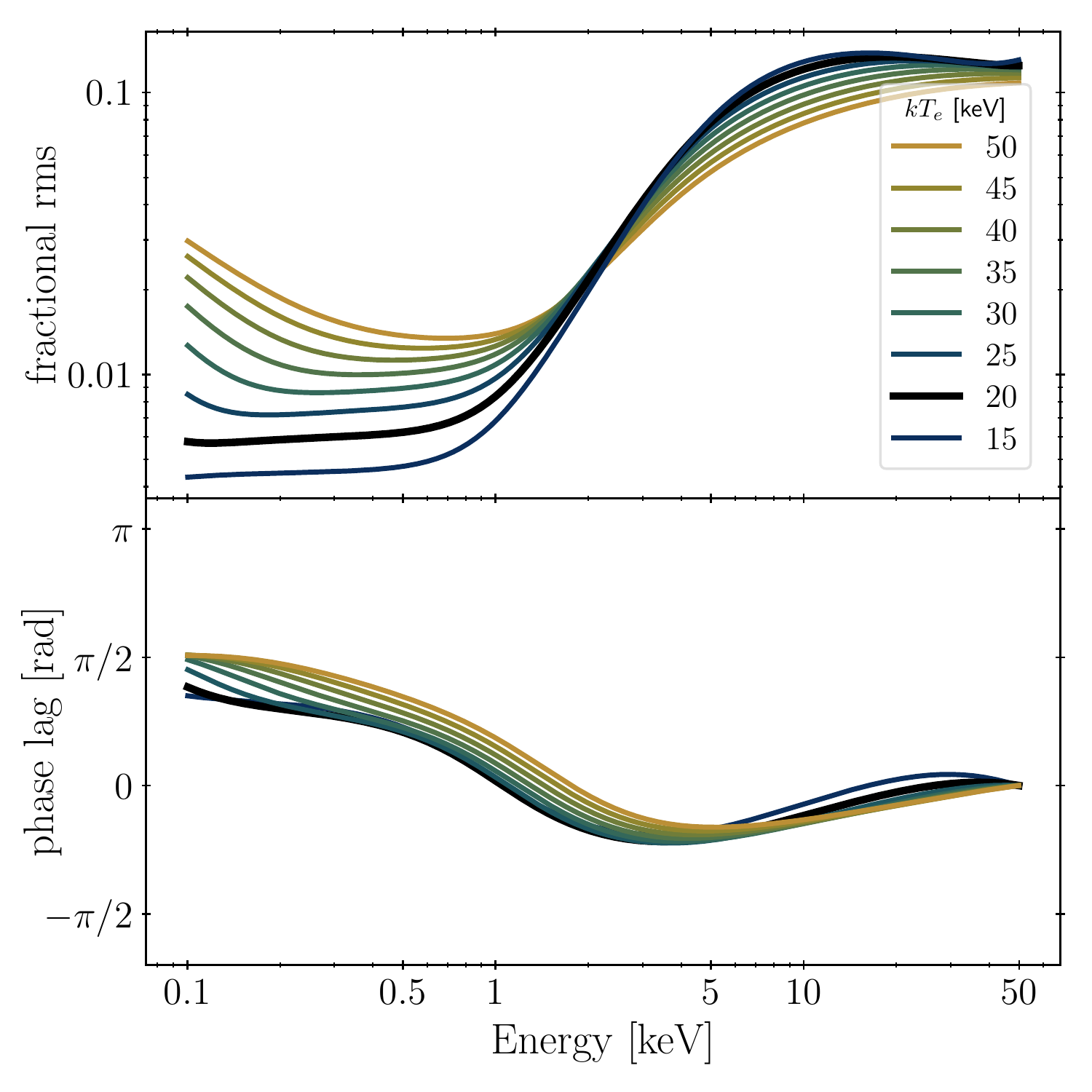}
    \,\,
    \includegraphics[width=\columnwidth]{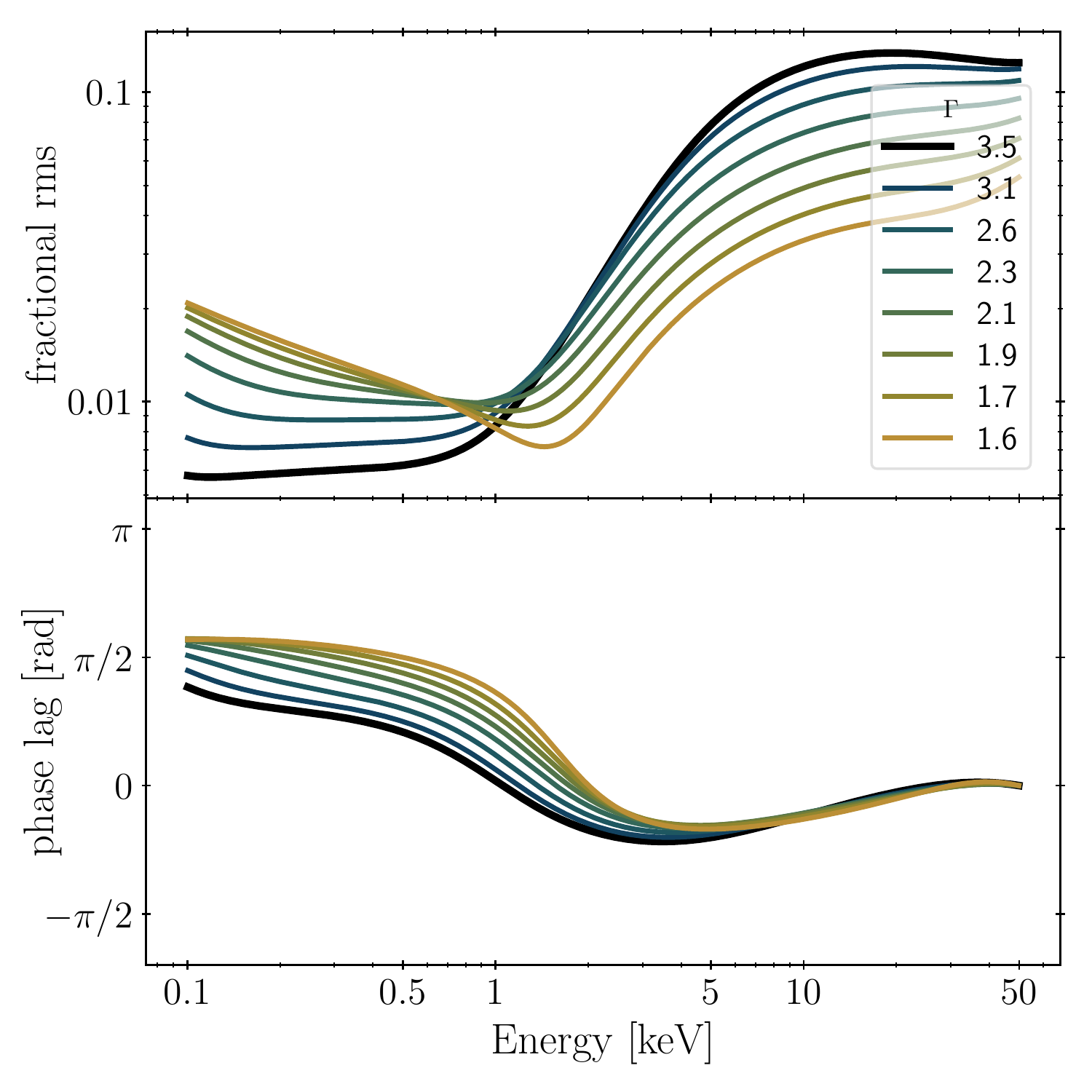}
    \caption{Same plot as Fig. \ref{fig:feedback_and_kTs} but here we explore the behaviour of the spectra varying independently the corona temperature and the photon index. On the left panel we consider $\eta=0.5$, $kT_s=0.5$~keV, $\Gamma=3.5$ and $kT_e$ assuming values from 15~keV (blue) to 50~keV (olive). On the right panel we fix $kT_e=20$~keV and we vary the photon index from 1.6 (olive) to 3.5 (blue). The black lines are the same as in Fig.~\ref{fig:feedback_and_kTs}}
    \label{fig:Tcorona_gamma}
\end{figure*}

\subsection{The effect of the corona parameters}

On the left panel of Fig.~\ref{fig:Tcorona_gamma} we show the rms and lag spectra as a function of the corona temperature, $kT_e$, for values from 15 keV (in blue) to 50~keV (in olive). Overall, the rms amplitude increases with energy from $\sim$1\% to $\sim$10\% becoming roughly flat above $\sim$10~keV. We see that at energies higher than $\gtrsim$2~keV, though very similar, the model predicts larger rms amplitudes for lower $kT_{\rm e}$ values. Around 2~keV all curves come together and as the energy decreases below $E \la 2$~keV, the trend of the rms amplitude with $kT_{\rm e}$ changes such that that the rms amplitude increases as $kT_{\rm e}$ increases. The lag spectra are smoother at high than at low values of $kT_e$. For energies below $\sim$3--5~keV the lags are soft, as they decrease with energy mainly due to feedback. For energies above $\sim$3--5~keV the lags become hard due to Comptonisation. For energies $E > kT_e$, the lags reach a local maximum and become soft again, due to Compton down-scattering.

The right panel of Fig.~\ref{fig:Tcorona_gamma} corresponds to different values of the photon index, $\Gamma$. As the electron temperature is fixed for every curve at $kT_e = 20$~keV, changes in the photon index reflect changes in the optical depth, $\tau$, following this relation: $\Gamma = \sqrt{\frac{9}{4} + \frac{1}{\frac{kT_e}{511~\mathrm{keV}}\tau(1 + \tau/3)}}~-~\frac{1}{2}$. We considered values of $\Gamma$ ranging from 1.6 in olive (which correspond to $\tau = 4.64$) to 3.5 in blue ($\tau = 1.3$). At energies below 1~keV, the model predicts higher rms amplitudes for lower $\Gamma$ values. In the 1--2~keV energy range a pivot point shows up for $\Gamma \lesssim 2.6$, and it becomes deeper as $\Gamma$ decreases. Meanwhile, above $\sim$2~keV the rms amplitudes are larger for higher values of $\Gamma$. For photons with energies below 4~keV the model yields softer lags for smaller photon indices. Concomitantly with the appearance of the pivot point in the rms, the fall in the lags becomes steeper for lower $\Gamma$, as the pivot point becomes deeper in the rms spectrum. Around 4~keV the curves start to overlap until it becomes difficult to separate one from the other for energies $\gtrsim$5~keV.

\subsection{Model application to \MAXI}

\MAXI~is a BH LMXB discovered in outburst on 2019 January 26 with the {\em MAXI} instrument on board the {\em ISS} \citep{2019ATel12425....1Y,2020ApJ...899L..20T}. The source showed a transition from the low-hard to the high-soft state, first detected with {\em MAXI} and later confirmed with {\em INTEGRAL} \citep{2019ATel12471....1C}, followed by a strong radio flare \citep{2019ATel12497....1C,2022arXiv220201514C}. Across the outburst, a thorough follow-up was performed with the \NICER instrument in the 0.2--12~keV energy range \citep{2012SPIE.8443E..13G}, and the spectral-timing analysis of a prominent type-B QPO observed during this transitions was presented in detail by \cite{2020MNRAS.496.4366B}. In their Figure~4, \cite{2020MNRAS.496.4366B} show the energy-dependent fractional-rms amplitude and phase lags of the type-B QPO in \MAXI. The strong QPO was detected in the full energy range, from 0.75 to 10~keV, at a stable QPO frequency $\nu_0 \approx 4.45$~Hz, with fractional rms amplitudes increasing from $\la$1\% at the lowest energies to $\sim$10\% at the highest ones. In addition, the lag-energy spectrum shows that photons with energies both below and above those in the 2--2.5~keV, the reference band, lag behind the photons in the reference band with an average delay of $\sim$32~ms for photons with energies below the reference band, and $\sim$21~ms for photons above that.

\cite{Garcia} applied the model originally developed by \cite{Karpouzas}, including a black-body source of soft photons, and found a relatively good fit to the lag spectrum, but failed to explain the high fractional rms values observed at high energies. They then introduced two independent, but physically coupled, Comptonisation regions, with which they achieved good fits to both the rms and lag spectra of the QPO. These two regions, which they called {\em small} and {\em large}. have sizes of $\sim$30 and $\sim$700~$R_g$, where $R_g$ is the gravitational radius assuming a 10~M$_\odot$ BH, with feedback fractions of $\sim$20 and $\sim$0.8\%, respectively. They also found a soft-photon source temperature of $kT_{\rm bb} \sim 0.35$~keV, lower than the temperature of the \diskbb~component from the best-fit to the time-averaged spectrum \citep[$\sim$0.6~keV,][]{2020MNRAS.496.4366B}; as explained by \cite{Garcia}, this is a consequence of the spectral differences between a \diskbb~and a \bb.

Given the use of a \diskbb~component in the model described here, we tried and fitted the \diskbb~variant of our variable-Comptonisation spectral-timing model, that we so-called \vkompthdk, to the same dataset.

\subsubsection{Single-component Comptonisation model \vkompthdk}

\begin{figure}
    \includegraphics[width=\columnwidth]{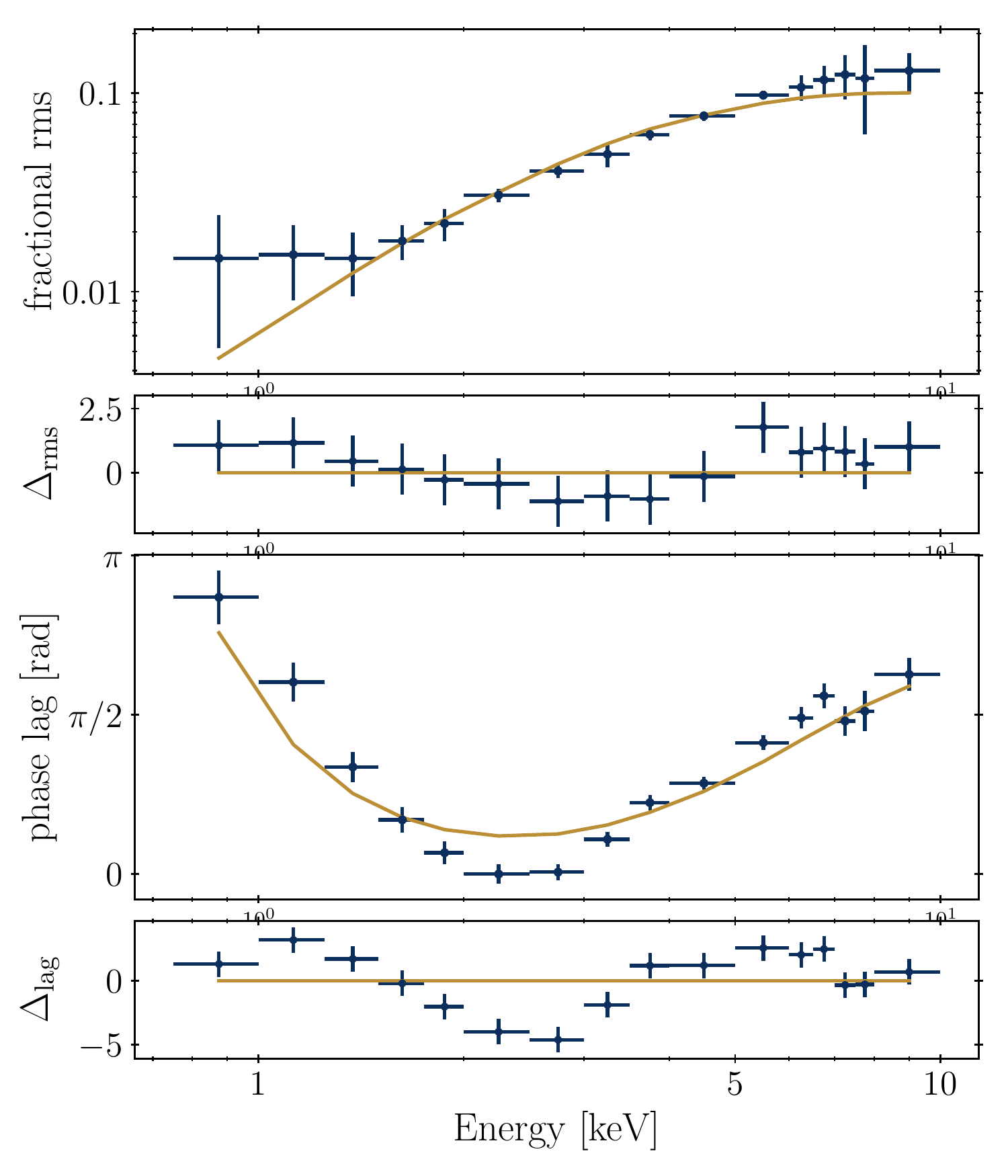}
    \caption{Fractional rms (upper panels) and phase-lag spectra (lower panels) of the 4.45~Hz type-B QPO of \MAXI. The solid lines represent the best-fitting \vkompthdk~model obtained for a corona size of $\sim$11\,000~km and a $\sim$35\% feedback fraction for a \diskbb~soft-photon source with a temperature of $\sim$0.45~keV. Residuals, $\Delta$\,=\,(data–model)/error, are also shown. The error bars of the rms amplitude were multiplied by a factor of five and rms dilution has been included (see text for details).}
    \label{fig:bestfit_vkompthdk}
\end{figure}

\begin{table*}
 \caption{Best-fitting values of the physical parameters of \MAXI~and their corresponding 1-$\sigma$ (68\%) uncertainties obtained with a single-component \vkompthdk~model.}
 \begin{tabular}{cccccccc}
  \hline
  $kT_e$ & $\Gamma$ & $\tau$ & $kT_s$ & $L$ & $\eta$ & ${\delta H}_{\rm ext}$ & $\chi_\nu^2$ (dof)\\
  (keV) & & & (keV) & ($10^3$ km) & & (\%) &  \\
  \hline
  20$^\dagger$ & 3.5$^\dagger$ & 1.3$^\dagger$ &  0.44$\pm$0.02 & 11$\pm$1 & 0.37$\pm$0.06 & 13$\pm$1 & 3.45 (27 dof) \\
  \hline
 \end{tabular}
 \flushleft{$^\dagger$ fixed parameters.}
 \label{tab:vkompthdk}
\end{table*}

In order to fit the fractional rms and phase lags of the type-B QPO of \MAXI~using the \vkompth~model introduced in this work, we used {\tt XSPEC v12.12.0g} \citep{1996ASPC..101...17A}. For this, we compiled the \vkompth~model written in Fortran against the {\tt XSPEC} libraries and loaded it as an external model; we also wrote the rms and phase lag spectra in PHA format and created the corresponding response (RMF) files. The best-fitting model was obtained using the Levenberg-Marquardt algorithm and uncertainties in the parameters were calculated by running Markov Chain Monte Carlo (MCMC) simulations of $10^5$ samples from 60 walkers, using the Goodman-Weare algorithm provided by {\tt XSPEC}. 

Along this process we kept the corona temperature, $kT_e$ fixed to 20~keV and the power-law index, $\Gamma$ to 3.5 \citep{2020MNRAS.496.4366B}, which yields an optical depth of $\sim$1.3\footnote{We note that the formula in \cite{Garcia} has a typo, but the values of the optical depth given in the paper are correct. The correct formula is $\tau = \sqrt{\frac{9}{4} + \frac{3}{\frac{kT_e}{511~\mathrm{keV}}\left[(\Gamma+1/2)^2-9/4\right]}}-\frac{3}{2}$.}. The rest of the physical parameters were left free to vary. As in \cite{Garcia}, while fitting the one-component \vkompthdk~model, we increased the errors in the fractional rms amplitudes by a factor of five to match the uncertainties in the lags. 

As our model only considers variations in the Comptonisation component, {\tt nthcomp}, whereas the best-fitting model to the time-average spectrum found by \cite{2020MNRAS.496.4366B} was {\tt diskbb+nthcomp+gaussian}, we introduced a correction to the fractional rms amplitude computed in the model to take into account the fraction of photons that are emitted by the accretion disk component {\tt diskbb} and the Fe-line {\tt gaussian}, and reach the observer without being Comptonised. This correction takes into account that, compared to the prediction of \vkompthdk, the observed fractional rms amplitudes are {\em diluted} by a factor ${\tt nthcomp}/{({\tt nthcomp+diskbb+gaussian})}$\footnote{We note that this assumes that the external \diskbb~and {\tt gaussian} components do not vary, at least not at the QPO frequency, and that all the variability comes from the {\tt nthcomp} component, which includes the \diskbb~soft-photons that are Comptonised in the corona.}.

In Figure~\ref{fig:bestfit_vkompthdk} we present the best-fitting model and residuals obtained with a single-Comptonisation region. The corresponding best-fitting parameters and their 1-$\sigma$ (68\%) uncertainties are summarized in Table~\ref{tab:vkompthdk}. Corner plots associated to the MCMC chains can be found in Appendix~\ref{appendix}. Similar to what was found by \cite{Garcia} using \vkompthbb, with a \bb~as the source of soft-photons, the \vkompthdk~model reproduces the trends of both the fractional rms amplitude and phase lags, but the fit is far from being statistically acceptable, with $\chi_\nu^2 = 3.45$ for 27 d.o.f. Using the \vkompthdk~model, the fit improves, yielding a higher fractional rms variability amplitudes than the \vkompthbb~model above $\sim$5~keV, as shown by the data, but fails to reproduce the lag spectrum at mid-energies ($2-3$~keV). The best-fitting model requires a relatively large corona of 11\,000~km ($\sim$600~$R_g$, for a 10~M$_\odot$ BH), with a relatively low feedback fraction of $\sim$35\%. The \diskbb~temperature is $\sim$0.45~keV, higher than the value found by \cite{Garcia} for a \bb~spectrum ($\sim$0.2~keV), but much more compatible with the best-fitting $kT_{\rm dbb}$ obtained by \cite{2020MNRAS.496.4366B} and \cite{2021MNRAS.505.3823Z} for the time-averaged spectra of the source, $kT_s\sim$0.6~keV. This result makes the model clearly more self-consistent in terms of the observed time-average and variable energy-dependent spectra. Finally, the amplitude of the oscillation of the external heating source is $\sim$13\%, higher than the value found by \cite{Garcia} for the fit with \vkompthbb\ ($\sim$4\%).

\subsubsection{Two-component Comptonisation model \vkdualdk}

\begin{figure}
    \includegraphics[width=\columnwidth]{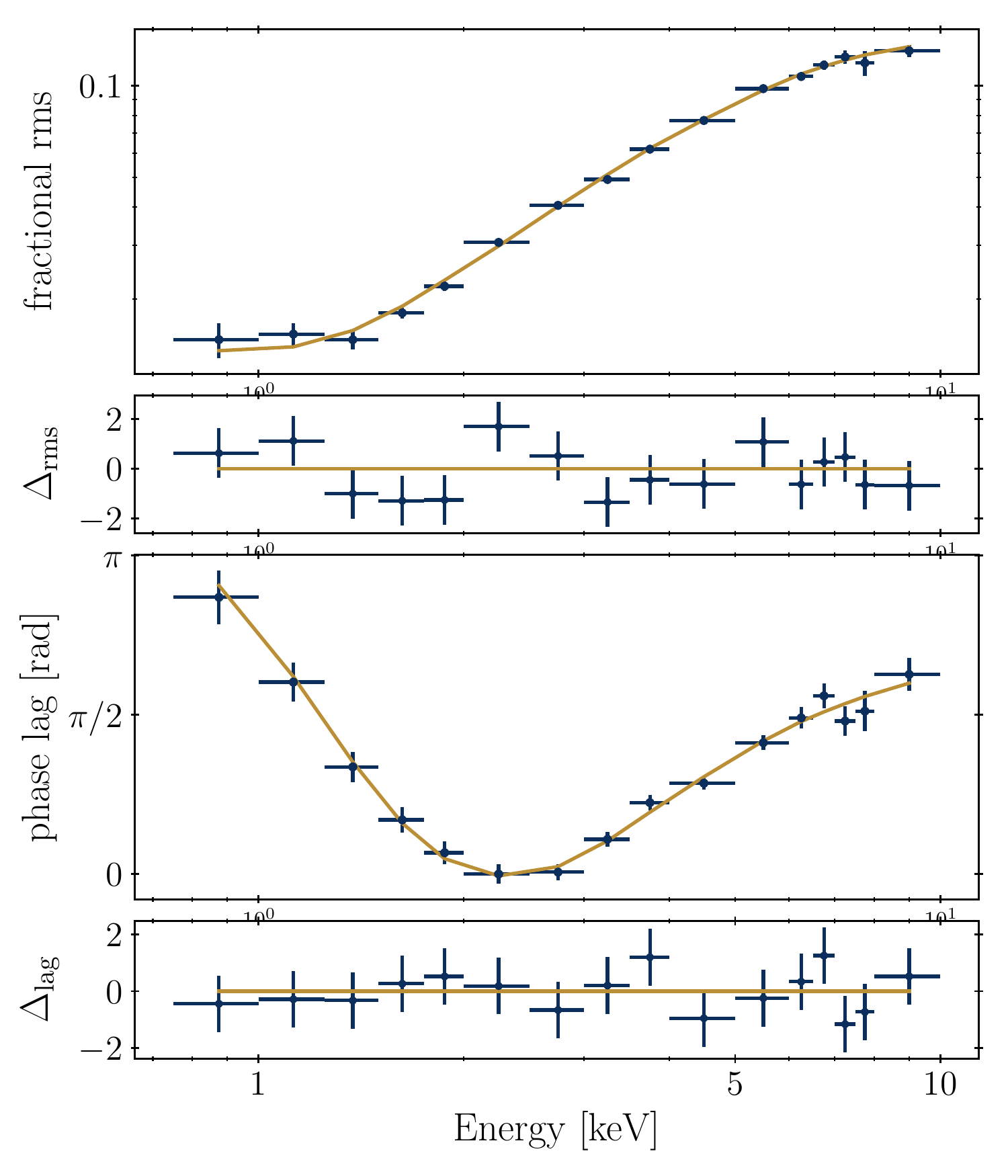}
    \caption{Energy-dependent fractional rms amplitude (upper panels) and phase lags (lower panels) of the 4.45 Hz type-B QPO of \MAXI. The best-fitting models (solid lines) are obtained using a combination of two Comptonisation components (\vkdualdk~model). Residuals, $\Delta$~=~(data–model)/error, are also shown. Dilution of the rms by the disk component has been included (see text for details).}
    \label{fig:bestfit_vkdualdk}
\end{figure}

\begin{table*}
 \caption{Same as Table~\ref{tab:vkompthdk} for the two-component \vkdualdk~model.}
 \begin{tabular}{ccccccccccccc}
  \hline
  $kT_e$ & $\Gamma$ & $\tau$ & $kT_{s,1}$ & $kT_{s,2}$ & $L_1$ & $L_2$ & $\eta_1$ & $\eta_2$ & ${\delta H}_{{\rm ext},1}$ & ${\delta H}_{{\rm ext},2}$ & $\phi$ & $\chi_\nu^2$ (dof)\\
  (keV) & & & (keV) & (keV) & ($10^3$ km) & ($10^3$ km) & & & (\%) & (\%) & (rad) & \\
  \hline
  20$^\dagger$ & 3.5$^\dagger$ & 1.3$^\dagger$ &  0.93$\pm$0.11 & 0.42$\pm$0.02 & 0.16$\pm$0.06 & 13.4$^{+3.6}_{-2.3}$ & >0.88 & 0.27$\pm$0.10 & 25$\pm$6 & 20$\pm$3 & 3.0$\pm$0.1 & 0.96 (22 dof) \\
  \hline
 \end{tabular}
 \flushleft{$^\dagger$ fixed parameters.}
 \label{tab:vkdualdk}
\end{table*}

Following \cite{Garcia}, and given the large $\chi^2$ obtained using the single-component Comptonisation model \vkompthdk, we also explored the possibility that the variability spectra arises from two Comptonisation regions that we also call {\em small (1)} and {\em large (2)}; we call this {\em dual} model \vkdualdk\footnote{We note that there was a typo in Eq.~3.1 of \cite{Garcia}. The minus sign should be a plus.}. For this, in order to match the properties of the time-averaged spectrum, we consider that both regions have $kT_{e, 1-2} = 20$~keV and $\Gamma_{1-2} = 3.5$ \citep[see][]{2020MNRAS.496.4366B,2021MNRAS.505.3823Z}. While fitting the QPO data with this \vkdualdk~model, we consider the actual 1-$\sigma$ errors of both rms and phase lags and take into account the rms {\em dilution} introduced in the previous sub-section. 

In Figure~\ref{fig:bestfit_vkdualdk} we show the best-fits and residuals found using the \vkdualdk~model. The corresponding best-fitting parameters and their 1-$\sigma$ (68\%) uncertainties are presented in Table~\ref{tab:vkdualdk}, where subindices 1 and 2 correspond to the {\em small} and {\em large} corona components, respectively. In  Appendix~\ref{appendix} we present the corner plot obtained from the MCMC chain for the most relevant physical parameters of this model. 

The best-fitting \vkdualdk~model matches the data statistically very well, with $\chi_\nu^2 = 0.96$ for 22 d.o.f, significantly better than the fit with single-component model. The dual model is able to explain simultaneously both the rms amplitudes leveling below $\sim$1.5~keV and the steady increase up to $\gtrsim$10\% at $\sim$5-10~keV, while showing the same overall trend of the $U$-shaped lag spectrum which also levels off in that same energy range. The residuals, on the other hand, show no systematic trends. The best-fitting model yields a {\em small} Comptonisation region of $\sim$150~km ($\sim$8~$R_g$ for a 10~M$_\odot$ BH) with very high feedback (>90\%) and a rather high seed temperature of $\sim$0.9~keV, pointing to the innermost parts of the \diskbb~component as the source of soft-photons, plus a {\em large} Comptonisation region of $\sim$12\,200~km ($\sim$650~$R_g$) with relatively low feedback fraction of $\sim$25\%, and a \diskbb~temperature of $\sim$0.4~keV, similar to the Comptonisation region found with the \vkompthdk~model presented in the previous sub-section. The oscillations of the two components have a lag, $\phi \sim$3.0$\pm$0.1~rad. The amplitude of the oscillation of the external-heating source are roughly $\sim$23\% and $\sim$20\%, respectively.

\subsection{Application to type-C QPOs in GRS~1915+105}

To verify whether this version of the model is also applicable to type-C QPOs in BH X-ray binaries, we fitted the \vkompthdk~model to the rms and lag spectra of the type-C QPO in the BH candidate GRS 1915+105 \citep{2020MNRAS.494.1375Z,2022mariano}. This type-C QPO was originally fitted using the \vkompthbb~model in 398 {\em RXTE} observations \citep{2022garcia}. In their Figures 4, 5 and 6 \cite{2022garcia} presented the best-fitting values of, respectively, the feedback fraction, $\eta$, the corona size, $L$, and the blackbody temperature, $kT_s$, as a function of the QPO frequency, $\nu_0$. Both $\eta$ and $kT_s$ show increasing trends with $\nu_0$, and two families of points are apparent in the plots: for $\nu_0 \lesssim 1.8$~Hz, the source shows low feedback fraction ($\eta \lesssim 0.3$) and low temperature ($kT_s \lesssim 0.2$~keV), whereas when $\nu_0 \gtrsim 2.5$~Hz the feedback becomes rather large ($\eta \gtrsim 0.8$) and the black-body temperature increases to $0.7-2$~keV. Meanwhile, in the intermediate $1.8-2.5$~Hz range, both families coexist. On the other hand, the corona size rapidly decreases from $\gtrsim$10$^4$~km to $\sim$10$^2$~km as $\nu_0$ increases from 0.5 to 2~Hz, and slowly increases again reaching intermediate sizes of $\sim$10$^3$~km when $\nu_0 \approx 4-6$~Hz . They also showed that, at the same QPO frequency, higher corona temperatures correspond to larger corona sizes.

When using the \vkompthdk~model presented in this work to fit this same QPO dataset, despite small differences in the individual values in each particular fit, for every relevant parameter we found consistent trends with those found in the earlier work from \cite{2022garcia} using \vkompthbb that were described above. This was somehow expected since, as we showed at the beginning of this Section, the differences between the \vkompthbb and \vkompthdk models become significant below $\sim$2~keV, while {\em RXTE} is only sensitive above 3~keV.

\section{Discussion} \label{discussion}

We developed a new version of the time-dependent Comptonization model of \citet{Karpouzas} in which we use an accretion disc instead of a blackbody as the source of seed photons that are subsequently inverse-Compton scattered in a corona. The new model can be used to fit the rms amplitude and phase lag spectra of low-frequency QPOs in BH-LMXBs. We also implemented a logarithmic grid to solve the time-dependent Kompaneets equation that describes the variability of the disk/corona system, which accelerates the code by two orders of magnitude compared to the version in \citet{Karpouzas}. 

The fractional rms variability of the LFQPOs in BH LMXBs increases with energy, from $\lessim$1\% at energies below 1-2~keV \citep{2020MNRAS.496.4366B} to 10-20\% at $\sim$30-50~keV 
\citep{2017ApJ...845..143Z,2020MNRAS.494.1375Z}, and in some cases even up-to 100-200~keV \citep[e.g.,][]{2018ApJ...866..122H,2021NatAs...5...94M}. At those energies, Comptonisation dominates the time-averaged spectrum and hence this process must be the mechanism responsible for the radiative properties of these QPOs. In our model, the QPO is a small oscillation of the source spectrum around the time-averaged spectrum, caused by fluctuations of the thermodynamic properties of either the Comptonising medium or the accretion disk. 

The model was originally developed for the high-frequency QPOs in NS-LMXBs considering a \bb spectrum, representing the surface of the NS itself or the boundary layer, as the source of soft-photons for Comptonisation \citep{LeeMiller,LeeMisra,KumarMisra14,Karpouzas}. The same model was later applied to model low-frequency QPOs in BH candidates \citep{Garcia,KostasTypeC}, who considered the same kind of soft-photon source. In this work we modified the model to have a disk-blackbody (\diskbb) as the source of soft-photons; the disc photons are injected into a spherical corona of homogeneously distributed hot electrons. The soft photons are inverse-Compton scattered by these electrons, and escape the corona with, on average, a higher energy than the one they had when they entered the corona. Because the photons that suffer more scatterings escape the corona at a later time, and with higher energies than those photons that escape directly or suffer less scatterings, the hard photons lag the soft ones (introducing hard lags). If a fraction of the up-scattered photons returns to the disc and is reprocessed and thermalised, a process that we call {\em feedback}, low-energy photons will escape the corona after the hard photons leading to soft lags. 

We note that our model is based upon two assumptions: (i) The soft-photon source \citep[a disk blackbody in this paper or a blackbody in][]{Karpouzas} is spherically symmetric and the seed photons are emitted isotropically in the corona \citep{SunyaevTitarchuk,1996MNRAS.283..193Z,1999MNRAS.309..561Z} (ii) The corona is spherically symmetric and homogeneous, and its optical depth, and therefore its density, remain constant during the oscillation. In the future we hope to generalize some of these hypotheses however, at this stage, these are necessary assumptions to be able to solve the mathematics involved in the problem in a semi-analytic way.

We found that when the soft-photon source is modelled as a \diskbb the fractional rms amplitude and phase lag spectra of the QPO have a smoother behaviour than when one uses a blackbody soft-photon source. In general, at high energies there is no significant difference in the rms and lag spectra between the \diskbb and blackbody cases, consistent with the fact that both time-averaged spectra, {\tt nthcomp} and {\tt bbody}, are almost identical at high energies. We also analyzed the dependence of both the rms amplitude and phase lags of the QPO upon the different physical parameters of the model. At energies above 1~keV, the rms amplitude depends upon the temperature of both the corona and the soft photon source, but primarily on the optical depth of the corona (Fig.~\ref{fig:Tcorona_gamma}), whereas the feedback fraction $\eta$ does not modify the rms spectrum. Instead, at energies below 1~keV, the rms amplitudes are affected by all parameters except for the soft-photon source temperature. The slope of the phase-lag spectrum, on the other hand, depends strongly on $\eta$. For a very low value of $\eta$ the lags are hard, while as $\eta$ increases the lags soften (see left panel of Fig.~\ref{fig:feedback_and_kTs}). While at energies above $\sim$5~keV the lag spectra are essentially independent of the parameter values, at energies below $\sim$5~keV the shape of the lag spectra changes significantly when the parameters change. For instance, the lags become softer with increasing corona temperature, $kT_e$, or with decreasing photon index, $\Gamma$, or feedback fraction, $\eta$.  Finally, we found a strong correlation between the minimum in the phase lag spectrum, the increasing energy in the rms, and the soft-photon source temperature, $kT_s$, which is apparent on the right panel of Fig.~\ref{fig:feedback_and_kTs}.

Our optimization of the numerical code, which now uses a logarithmic energy grid, makes it fast enough such that it is computationally possible to use our \vkompth~model\footnote{\url{https://github.com/candebellavita/vkompth}} as a local model of {\tt XSPEC} to fit the time-dependent as well as the time-averaged spectra in the standard way. As an example of the application of the model to a low-frequency QPO in a BH LMXB, we have chosen to fit the available data from the type-B in \MAXI obtained while the source was transitioning from the high to the soft states \citep{2020MNRAS.496.4366B}. This dataset was early fitted using our model with a \bb soft-photon source by \cite{Garcia}. The available data comprises the energy-dependent fractional rms and phase lags associated to this QPO with a $\nu_0 = 4.55$~Hz.  When comparing those fits obtained using a \diskbb~model to those from \cite{Garcia} based on a \bb~source of soft-photons, we found that the former are statistically better, and that the physical parameters retrieved are more appropriate when compared both to those derived from the steady-state or time-averaged spectra, and to the overall scenario.

Following \cite{Garcia}, we initially fitted simultaneously the rms and lag spectra of this QPO with the single-Comptonisation region model, \vkompthdk in {\tt XSPEC}. The best-fitting model reproduces the trends in the data quite well, but the fit is statistically unacceptable, with a high reduced $\chi_\nu^2 = 3.45$. The fit requires a large corona, $L \sim 600~R_g$, with relatively low feedback, $\eta \sim 35$\%. While this may appear to be too large a corona, we note that such a large corona ($L \sim 100-500~R_g$) was also found in MAXI~J1820+70 during the transition from the high to the soft state, from fits to the broadband lags using a lamp-post reverberation model \citep{2021ApJ...910L...3W}, and a large corona, $L\sim$100~$R_g$, was also inferred in the case of Cyg X-1 using polarimetric measurements from PoGO+ \citep{2018NatAs...2..652C}. We also fitted a model with two-Comptonisation regions, which we call \vkdualdk, to the same data. We found that the best-fitting model requires a {\em large} corona with similar characteristics to those found using the \vkompthdk~model, and a {\em small} corona of $L\sim$150~km ($\sim$8~$R_g$ for a 10~M$_\odot$ BH) with a very high feedback fraction, $\eta >$90\%. 
We interpret this solution as a {\em dual} corona formed by a compact and hot region situated very close to the BH and fed by the innermost parts of the accretion disk with a relatively high effective temperature, $kT_s \sim 0.9$~keV, and a large and extended region fed by a more external part of the accretion disk, and thus having a lower effective temperature, $kT_s\sim$0.4~keV); this large region may be associated to the base of the extended jet which starts to develop during the state transition in \MAXI~\citep{2022arXiv220201514C}. 

We also fitted the \vkompthdk model presented here to 398 \RXTE observations of GRS~1915+105 containing a type-C QPO. For every relevant physical parameter of the model, we found consistent trends with QPO frequency to those obtained using a black body, \vkompthbb, as the soft-photon source \citep[see Figures 4, 5 and 6 in][]{2022garcia}. This was expected given the energy coverage from \RXTE, which was not sensitive below $\sim$3~keV. In addition, recently \citet{Zhang2022} successfully applied the \vkompthdk~model to fit the type-C QPOs in MAXI J1535--571, using {\em HXMT} data ranging from 1 to 100~keV. Therefore, we conclude that this variable-Comptonisation model is able to explain both type-B and type-C low-frequency QPOs in BH LMXB binaries.

\section*{Acknowledgements}

We are grateful to an anonymous referee for constructive comments that helped us improve this paper. CB is a fellow of the Consejo Interuniversitario Nacional, Argentina. FG is a CONICET researcher. FG acknowledges support by PIP 0113 (CONICET). This work received financial support from PICT-2017-2865 (ANPCyT). This work is part of the research programme Athena with project number 184.034.002, which is (partly) financed by the Dutch Research Council (NWO). CB thanks Eduardo M. Guti\'errez for valuable help with the numerical code. FG and MM acknowledge fruitful discussions held at the ISSI Bern meeting.

\section*{Data Availability}

 The model developed in this paper is available through \url{https://github.com/candebellavita/vkompth}. We encourage people interested in applying the model to data to contact the authors for further support.
 



\bibliographystyle{mnras}
\bibliography{paper} 




\appendix

\section{MCMC corner plots}
\label{appendix}

\begin{figure*}
    \includegraphics{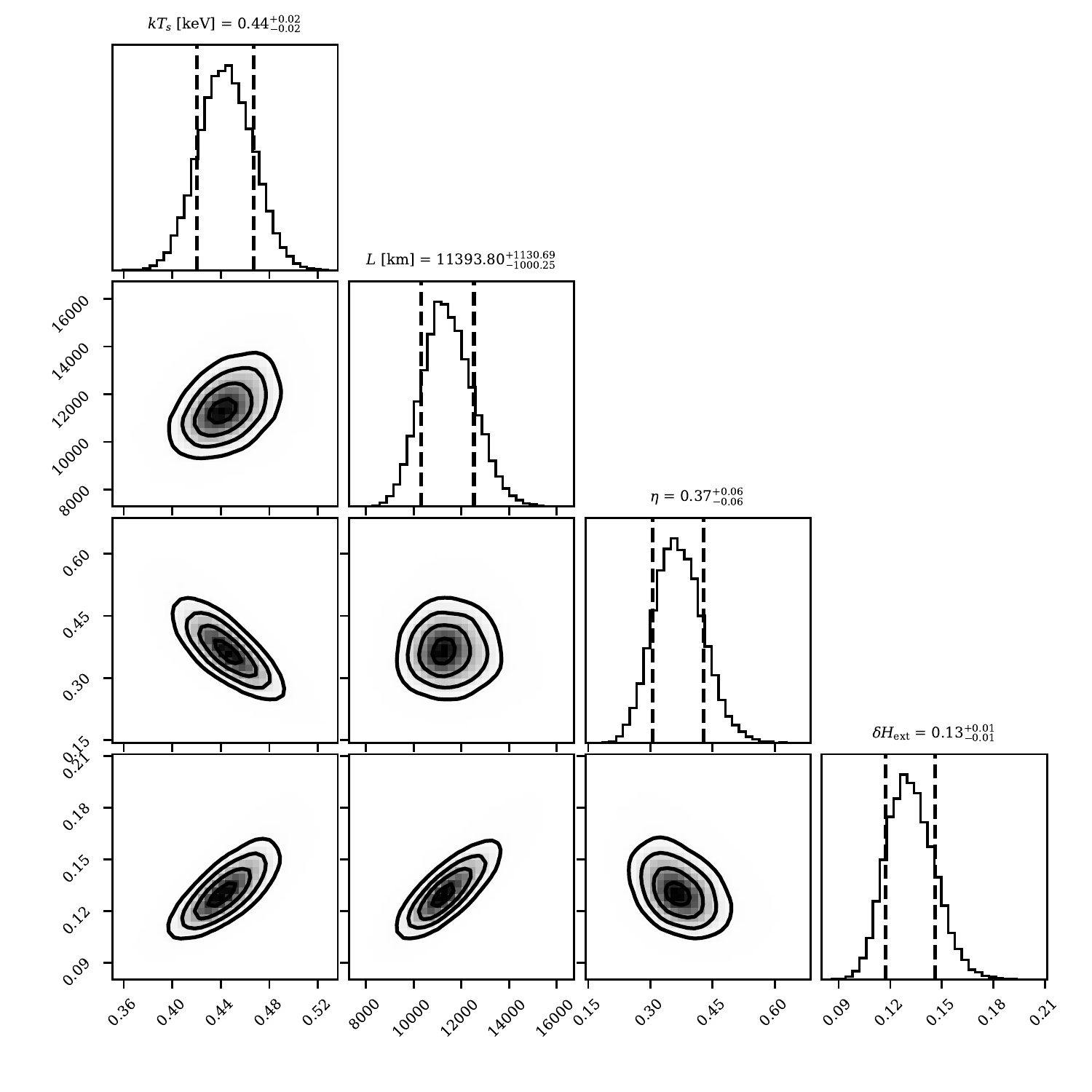}
    \caption{Corner plot of the physical parameters of \MAXI~ obtained with the single Comptonisation model \vkompthdk~introduced in this paper, which includes a \diskbb~soft-photon source characterised by its temperature, $kT_s$. The parameters of the corona include its physical size, $L$, the feedback fraction, $\eta$, and the amplitude of the external heating rate oscillation, $\delta H_{\rm ext}$. Median and 68\% credible-interval for each parameter are indicated at the top of each diagonal panel.}
    \label{fig:corner_vkompthdk}
\end{figure*}

\begin{figure*}
    \includegraphics{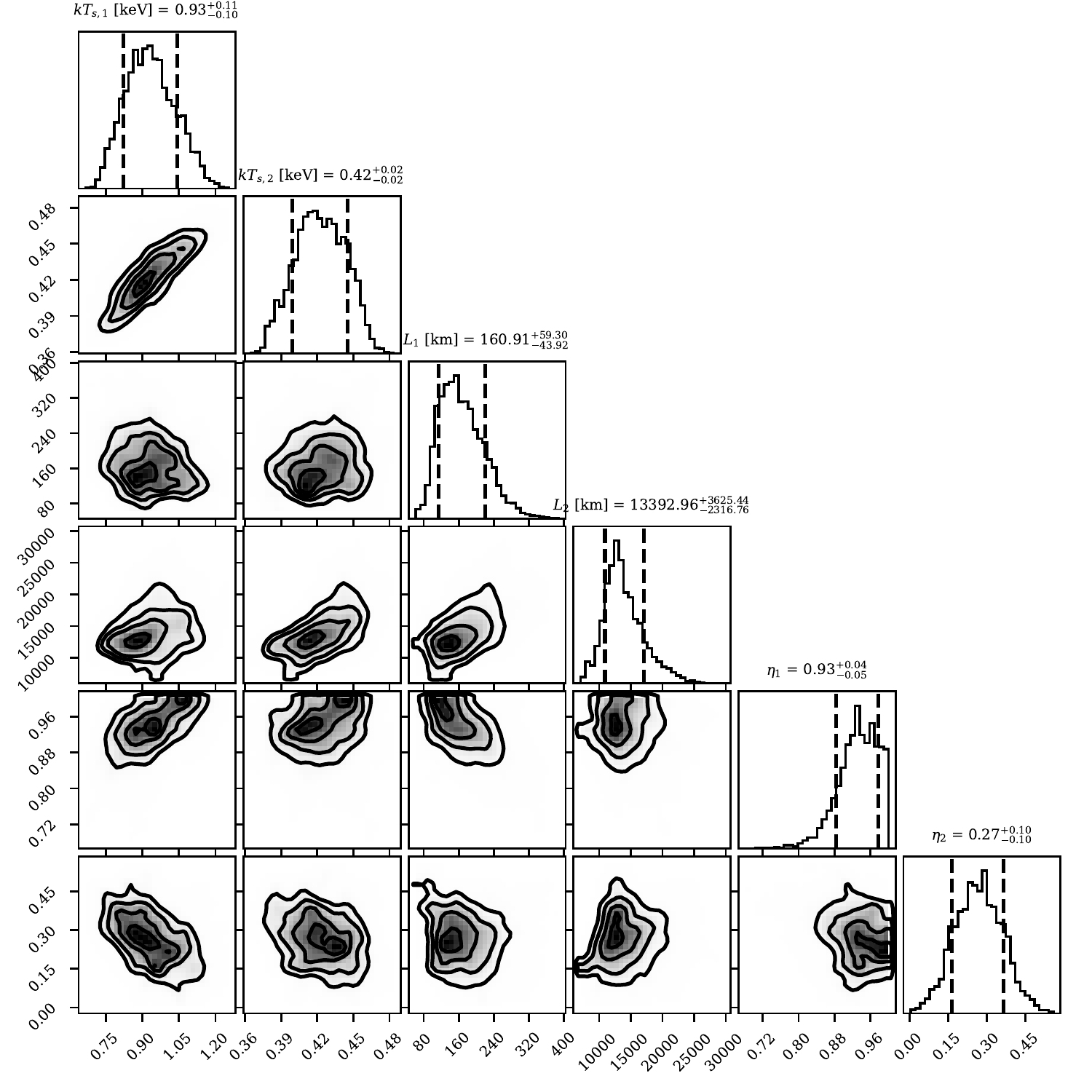}
    \caption{Same as Figure~\ref{fig:corner_vkompthdk} but for the main best-fitting parameters obtained using a two-component {\em dual} model, named \vkdualdk.}
    \label{fig:corner_vkdualdk}
\end{figure*}

As explained in Sec.~\ref{results}, we fitted the \vkompth~model using {\tt XSPEC} \citep{1996ASPC..101...17A} using the Levenberg-Marquardt algorithm. After finding the appropriate minimum, in order to estimate the uncertainties in the physical parameters provided by the models, we ran MCMC chains of $10^5$ steps using 60 walkers under the Goodman-Weare scheme in {\tt XSPEC}. We then used the {\tt tkXspecCorner}\footnote{\url{https://github.com/garciafederico/pyXspecCorner}} program to interactively create corner plots of the main physical parameters that we present on Figures~\ref{fig:corner_vkompthdk}~and~\ref{fig:corner_vkdualdk} for the \vkompthdk~and the \vkdualdk~models, respectively. As usual, in the lower-left panels of the corner plots we show the joint-probabilities of each pair of relevant physical parameters of the models, while in the diagonal we show the individual probability of each parameter, including both the median and 68\% confidence intervals in the top labels. 

\bsp	
\label{lastpage}
\end{document}